\documentstyle[aps,prl,epsf,floats]{revtex}

\def\PRL #1 #2 #3 {Phys.\ Rev.\ Lett.\ {\bf #1}, #2 (#3)}
\def\PRD #1 #2 #3 {Phys.\ Rev.\ D~{\bf #1}, #2 (#3)}
\def\PLB #1 #2 #3 {Phys.\ Lett.\ B~{\bf #1}, #2 (#3)}
\def\NPB #1 #2 #3 {Nucl.\ Phys.\ {\bf B#1}, #2 (#3)}
\def\ZPC #1 #2 #3 {Z.\ Phys.\ C~{\bf #1}, #2 (#3)}

\newcommand{\gtap}{\;{\raise.3ex\hbox{$>$\kern-.75em\lower1ex\hbox{$\sim$}}}\;}
\newcommand{\ltap}{\;{\raise.3ex\hbox{$<$\kern-.75em\lower1ex\hbox{$\sim$}}}\;}

\begin{document}

\draft

\title{Upper bound on the scale of Majorana-neutrino mass generation}
\author{F.~Maltoni, J.~M.~Niczyporuk, and S.~Willenbrock}
\address{Department of Physics, University of Illinois at Urbana-Champaign,
Urbana, IL\ \ 61801}
\date{January 18, 2001}

\maketitle

\begin{abstract}
We derive a model-independent upper bound on the scale of Majorana-neutrino
mass generation.  The upper bound is
$4\pi v^2/\sqrt 3 m_\nu$, where $v \simeq 246$ GeV is the weak
scale and $m_\nu$ is the Majorana neutrino mass.  For neutrino masses
implied by neutrino oscillation experiments, all but one of these
bounds are less than the Planck scale, and they are all within a few orders
of magnitude of the grand-unification scale.
\end{abstract}


There are three known types of neutrinos in nature, associated with the 
electron, the muon, and the tau lepton.  Considerable evidence
has mounted that one or more of these neutrino species has a nonzero mass,
based on the observation of neutrino oscillations \cite{Rob}.
Since neutrinos are massless in the standard model of particle physics, the observation of nonzero neutrino masses is our 
first evidence of physics beyond the standard model.

The standard model of the electroweak interaction is a 
gauge theory based on the local symmetry group 
SU(2)$_L\times$U(1)$_Y$. The model contains three generations
of quark and lepton fields and an SU(2)$_L$-doublet Higgs field which acquires
a vacuum-expectation value and breaks the SU(2)$_L\times$U(1)$_Y$ symmetry to 
the U(1)$_{EM}$ symmetry of electromagnetism.  There are three 
reasons why neutrinos are massless in this model:
\begin{enumerate}

\item Only renormalizable interactions are included, {\it i.e.}, terms 
in the Lagrangian of mass dimension four or less.  The unique term of 
dimension five allowed by the gauge symmetry is \cite{W}
\begin{equation}
{\cal L} =\frac{c}{M}(L^T\epsilon\phi) C (\phi^T\epsilon L)+h.c.\;,
\label{DIM5}
\end{equation}
where $L=(\nu_L,\ell_L)$
is an SU(2)$_L$-doublet containing the left-chiral neutrino
and charged-lepton fields and $\phi=(\phi^+,\phi^0)$ is the SU(2)$_L$-doublet 
Higgs field ($\epsilon \equiv i \sigma_2$ is the 
antisymmetric $2\times 2$ matrix in SU(2)$_L$ space;  $C$ is the 
charge-conjugation matrix in Dirac space).
This term would give rise to a Majorana
neutrino mass $m_\nu = cv^2/M$ when the neutral component of the Higgs
field acquires a vacuum-expectation value $\langle\phi^0\rangle = v/\sqrt 2$,
where $v = (\sqrt 2 G_F)^{-1/2} \simeq 246$ GeV is the weak scale.

\item The only neutral lepton fields are in SU(2)$_L$ doublets.
In particular, no SU(2)$_L\times$U(1)$_Y$-singlet fermion field is
present.  If present, the gauge symmetry would allow a Yukawa term
\begin{equation}
{\cal L} = -y_D \bar L\epsilon \phi^*\nu_R+h.c.\;, 
\label{DIRAC}
\end{equation}
where $\nu_R$ is the singlet field.  Such a term would 
result in a Dirac neutrino
mass $m_D = y_D v/\sqrt 2$ when the neutral component of the Higgs
field acquires a vacuum-expectation value, in the same
way that the other fermions acquire Dirac masses.  The gauge symmetry
would also allow a Majorana mass for the singlet field,
\begin{equation}
{\cal L} = -\frac{1}{2}M_R \nu_R^T C \nu_R+h.c.\;.
\label{MAJORANA}
\end{equation}
Majorana neutrino masses may also be generated via the addition of
SU(2)$_L$-triplet, $Y=0$ fermion fields \cite{JS}.

\item The only scalar field is the SU(2)$_L$-doublet Higgs field.
In particular, no SU(2)$_L$-triplet, $Y=1$ Higgs field is present. If present,
the gauge symmetry would allow a term 
\begin{equation}
{\cal L} = -y_M L^T\epsilon\sigma^iCL\Phi^i+h.c.\;,
\label{TRIPLET}
\end{equation}
where $\Phi^i$ is the Higgs triplet field.  Such a term would result in
a Majorana neutrino mass $m_\nu = 2y_M u$ when the neutral component 
of the Higgs triplet field, $\Phi^0 = (\Phi^1+i\Phi^2)/\sqrt 2$, 
acquires a vacuum-expectation value $\langle\Phi^0\rangle=u/\sqrt 2$ 
\cite{CL,GR,GGN,MS}.
Majorana neutrino masses may also be generated via the addition of
SU(2)$_L$-singlet scalar fields \cite{CL,Z}.
\end{enumerate}
These restrictions eliminate the possibility of a Dirac neutrino mass
and yield an ``accidental'' global lepton-number symmetry,
U(1)$_L$, which forbids a Majorana neutrino mass.\footnote{Lepton
number guarantees masslessness of the neutrino to all orders
in perturbation theory.  Beyond perturbation theory, lepton number is
violated; however, $B-L$ symmetry 
(baryon number minus lepton number) survives and suffices to enforce the
masslessness of the neutrino \cite{GGN}.}  
In this paper, we will encounter examples with massive neutrinos based on 
relaxing each of these three restrictions.\footnote{In the minimal 
supersymmetric standard model, renormalizable Majorana-neutrino mass terms
are allowed.  Imposing $R$-parity suffices to forbid such terms.}

Since neutrino masses are necessarily associated with physics beyond the 
standard model, one would like to know the energy scale at which this new
physics resides.  In this paper we derive a model-independent 
upper bound on the scale
of Majorana-neutrino mass generation.  We also discuss two models that
exemplify, and can even saturate, this bound: one with an 
SU(2)$_L\times$U(1)$_Y$-singlet fermion field,
and one with an SU(2)$_L$-triplet Higgs field.  The analysis we perform
is in the spirit of a similar analysis for Dirac fermions carried out
in Ref.~\cite{AC}.  However, there is no known model that saturates the 
upper bound on the scale of Dirac-fermion mass generation \cite{AC,Go},
in contrast to the case of Majorana neutrino masses addressed in this paper.

We assume that the neutrino masses are Majorana, unlike the other known 
fermions, which carry electric charge and are therefore forbidden to have 
Majorana masses.  If there is no 
SU(2)$_L\times$U(1)$_Y$-singlet fermion field in nature, then the neutrino
masses are necessarily Majorana.  
However, even if such a field exists, the gauge
symmetry allows the  Majorana mass term of Eq.~(\ref{MAJORANA}) for this field,
and there is no reason why this mass should be small.  The other
known fermions acquire a mass only after the SU(2)$_L\times$U(1)$_Y$
symmetry is broken, and thus their masses are of order the weak scale,
$v$, or less.  
Since a Majorana mass for the $\nu_R$ field is not
protected by the gauge symmetry, it is natural to assume that it would be 
much greater than the weak scale \cite{G}.  
So even if the $\nu_R$ field exists it
is likely to be heavy, in which case the light neutrinos are Majorana fermions.

We begin our analysis with the standard model, but with a Majorana neutrino 
mass of unspecified origin.  Since the neutrino mass is put in artificially, 
this is only an effective field theory, valid up to some energy scale 
at which it is subsumed by
a deeper theory, which we regard as the scale of Majorana-neutrino mass
generation.  The effective theory yields amplitudes that are an expansion
in powers of energy divided by some mass scale.  
A simple way to derive an upper bound on the scale at which 
the effective theory breaks down is to examine tree-level $2\to 2$
scattering amplitudes and identify the 
ones that grow with energy.  Unitarity of the $S$-matrix ensures that 
partial-wave amplitudes of inelastic $2\to 2$ scattering processes 
cannot exceed $1/2$ in absolute value.  
When that value is exceeded at tree level,
it indicates that the effective field theory is no longer valid, because
the energy expansion does not converge.  We thereby
discover the energy at which the effective field theory necessarily breaks 
down; this represents an upper bound on the scale of new physics.  This
argument has been used to derive an upper bound on the scale of new 
physics in the Fermi theory of the weak interaction \cite{LY}, on the scale of 
electroweak symmetry breaking in the electroweak theory (without a Higgs
field) \cite{CG}, and on the scale of Dirac-fermion 
mass generation \cite{AC}.

The scattering amplitudes that grow with energy involve Majorana neutrinos
in the initial and/or intermediate state, and longitudinally-polarized 
weak vector bosons in the final state.  
The Feynman diagrams that contribute to the four relevant
amplitudes are shown in Fig.~1.  In the high-energy limit, $s \gg M_W^2, M_Z^2,
m_\nu^2, m_\ell^2$, the zeroth-partial-wave amplitudes are given by the 
simple expressions
\begin{eqnarray}
a_0\left(\frac{1}{\sqrt 2}\nu_{i\pm} \nu_{j\pm} \to W^+_LW^-_L\right) & 
\sim & \mp\frac{m_{\nu_i} \sqrt s}{8\pi\sqrt 2v^2}\delta_{ij} \label{WW}\\ 
a_0\left(\frac{1}{\sqrt 2}\nu_{i\pm} \nu_{j\pm} 
\to \frac{1}{\sqrt 2}Z^0_LZ^0_L\right) & 
\sim & \mp\frac{m_{\nu_i} \sqrt s}{8\pi v^2}\delta_{ij} \label{ZZ}\\ 
a_0(\nu_{i-} \ell_- \to Z^0_LW^-_L) & 
\sim & \frac{m_{\nu_i} \sqrt s}{8\pi\sqrt 2v^2}U^*_{\ell i} \label{ZW}\\ 
a_0\left(\frac{1}{\sqrt 2}\ell_- \ell_- 
\to \frac{1}{\sqrt 2}W^-_LW^-_L\right) & 
\sim & \frac{\sqrt s}{8\pi v^2}\sum_{i=1}^3 
U_{\ell i}^2 m_{\nu_i} \label{SS}
\end{eqnarray}
where $v$ is the weak scale, the indices $i,j$ denote the 
three neutrino mass eigenstates, 
the subscripts on the neutrinos and charged leptons indicate 
helicity $\pm 1/2$, and the subscript on the partial-wave 
amplitude indicates $J=0$.
The unitary matrix $U_{\ell i}$ relates the neutrino weak and mass 
eigenstates.
Each amplitude grows linearly with energy, and is proportional to the Majorana
neutrino mass or a linear combination of masses.\footnote{The amplitude for 
$\ell_- \ell_- \to W^-_LW^-_L$ involves the same linear 
combination of masses as the amplitude for neutrinoless double beta 
decay \cite{Ri}.}

\begin{figure}[t]
\begin{center}
\vspace*{0cm}
\hspace*{0cm}
\epsfxsize=8.6cm \epsfbox{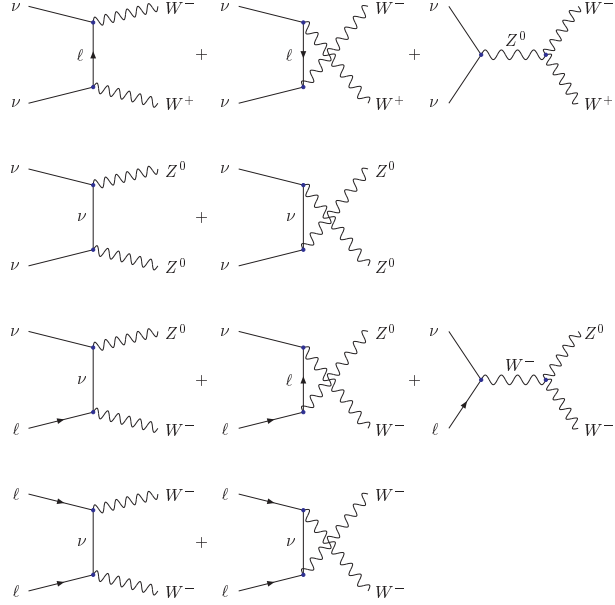}
\vspace*{0cm}
\caption{Feynman diagrams that contribute to the amplitudes in 
Eqs.~(\ref{WW})--(\ref{SS}).  The source of the Majorana neutrino mass
is unspecified, so there are no diagrams involving the coupling of the
Majorana neutrino to the Higgs boson.  Unitary gauge is used throughout.}
\label{fig:fig1}
\end{center}
\end{figure}

The strongest bound on the scale of Majorana-neutrino mass generation is
obtained by considering a scattering process which is a linear combination
of the above amplitudes:
\begin{equation}
a_0\left(\frac{1}{2}(\nu_{i+}\nu_{i+} - \nu_{i-}\nu_{i-}) \to 
\frac{1}{\sqrt 3}(W^+_LW^-_L + Z^0_LZ^0_L)\right)
\sim -\frac{\sqrt 3m_{\nu_i} \sqrt s}{8\pi v^2}\;.
\end{equation}
The unitarity condition on inelastic $2\to 2$ scattering amplitudes,
$|a_J| \leq 1/2$ \cite{MVW}, implies that the scale of Majorana-neutrino
mass generation is less than the scale
\begin{equation}
\Lambda_{Maj} \equiv \frac{4\pi v^2}{\sqrt 3 m_{\nu}}\;,
\label{BOUND}
\end{equation}
which is inversely proportional to the neutrino mass.  This is the principal
result of this paper.

To gain some intuition for Eq.~(\ref{BOUND}), 
we consider three different mechanisms
for the generation of a Majorana neutrino mass.  First consider the 
addition of the dimension-five term of Eq.~(\ref{DIM5}) to the standard model.
The neutrino acquires a Majorana mass $m_\nu = cv^2/M$, where $c/M$ is the
coefficient of the dimension-five term.  However,
despite the addition of an explicit source for the Majorana
neutrino mass, the theory remains an effective field theory.  The 
dimension-five term generates a $\nu\nu h^0$ vertex, where $h^0$ is the 
Higgs boson, which leads to the additional Feynman diagram in Fig.~2.
Although this diagram cancels the term that grows with energy in
the amplitude of  
Eq.~(\ref{ZZ}), the other three amplitudes continue to grow with 
energy.\footnote{The amplitude of Eq.~(\ref{WW}) undergoes a sign change
when the Higgs diagram is included.  The other two amplitudes have no
additional contributions.}  Thus there must still be new physics at or
below the scale $\Lambda_{Maj}$.  The generation of a Majorana neutrino mass
via a nonrenormalizable dimension-five term cannot promote an 
effective field theory to a renormalizable one.

\begin{figure}[t]
\begin{center}
\vspace*{0cm}
\hspace*{0cm}
\epsfxsize=3.2cm \epsfbox{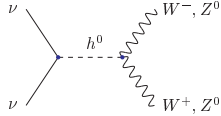}
\vspace*{0cm}
\caption{Additional diagram that contributes to the amplitudes in 
Eqs.~(\ref{WW}) and (\ref{ZZ}) when the Majorana neutrino acquires its
mass via a coupling to the Higgs field.}
\label{fig:fig2}
\end{center}
\end{figure}

Consider instead the addition of an SU(2)$_L\times$U(1)$_Y$-singlet fermion
field, $\nu_R$, and the terms in Eqs.~(\ref{DIRAC}) and (\ref{MAJORANA}),
which are allowed by the gauge symmetry.  Let us consider the limit
$M_R \gg m_D$, motivated by our earlier argument that $M_R$ should be much
greater than the weak scale while the Dirac mass $m_D$ is of order or less 
than the weak scale.  This ``see-saw'' model yields a Majorana 
neutrino of mass $m_\nu \approx m_D^2/M_R$, which is much less than the 
Dirac mass, and thus provides an attractive explanation of why neutrinos
are so much lighter than the other known fermions \cite{GRS}.  There is also
a heavy Majorana neutrino, $N \approx \nu_R$, approximately of mass $M_R$.  
This particle
leads to the additional Feynman diagrams obtained by replacing any intermediate
$\nu$ state in Fig.~1 with $N$.  A Higgs diagram analogous to 
Fig.~2 must also be included.  One finds that the terms that grow with 
energy are cancelled in
all four amplitudes of Eqs.~(\ref{WW})--(\ref{SS}), so the addition of 
these two dimension-four terms has promoted the effective field theory
to a renormalizable one.  The scale of Majorana-neutrino mass generation
is the mass of the heavy Majorana neutrino, 
$M_R \approx m_D^2/m_\nu$, and since $m_D = y_Dv/\sqrt 2$, one finds 
$M_R \approx y_D^2 v^2/2m_\nu$.  This respects the upper bound
of Eq.~(\ref{BOUND}) provided the Yukawa coupling 
$y_D \ltap \sqrt{8\pi}$, as it must \cite{CFH}.  The bound
is saturated when the Yukawa coupling takes its largest allowed value.

The third mechanism introduces an SU(2)$_L$-triplet Higgs field and the term
of Eq.~(\ref{TRIPLET}) to generate a Majorana neutrino 
mass \cite{CL,GR,GGN,MS}.  The vacuum-expectation value of this field must 
be much less than the weak scale, because the relation 
$M_W^2 \simeq M_Z^2\cos^2\theta_W$, which is satisfied experimentally, 
is obtained if the weak bosons acquire their mass dominantly from the 
vacuum-expectation value
of an SU(2)$_L$ doublet, but not a triplet.  In any case, 
a small vacuum-expectation
value for the triplet is desirable in order to generate small Majorana
neutrino masses ($m_\nu = 2y_Mu$).
This model contains three neutral scalars, one singly-charged scalar, and
one doubly-charged scalar.  The term of Eq.~(\ref{TRIPLET}) gives rise to new
interactions that yield the additional Feynman diagrams in Fig.~3 
involving these Higgs scalars in the intermediate state.\footnote{We impose
CP conservation in this model, in which case one of the neutral scalars is
CP odd and does not contribute to the amplitudes.}  These diagrams
cancel the terms that grow with energy in the amplitudes of 
Eqs.~(\ref{WW})--(\ref{SS}),
so once again the addition of a dimension-four term has rendered an 
effective field theory renormalizable.  The scale of Majorana-neutrino 
mass generation
is the mass of these Higgs scalars.  
We have shown that their mass
respects, and can even saturate, the model-independent upper bound on
the scale of Majorana-neutrino mass generation, Eq.~(\ref{BOUND}).

These two models demonstrate that $M$, the inverse coefficient of
the dimension-five term of Eq.~(\ref{DIM5}), is the scale of 
Majorana-neutrino mass generation.  If one integrates out the heavy 
Majorana neutrino in the see-saw model, one obtains this dimension-five
term, with $c/M=-y_D^2/2M_R$.  The same thing happens if one 
integrates out the Higgs triplet. 
In both cases, $M$ is equal to the scale at which new
physics appears, and $c$ is a dimensionless product of coupling constants and
mass ratios.  These models can naturally saturate our
bound, Eq.~(\ref{BOUND}), precisely because they generate a Majorana-neutrino
mass term of dimension five in the low-energy theory.

\begin{figure}[h]
\begin{center}
\vspace*{0cm}
\hspace*{0cm}
\epsfxsize=8.6cm \epsfbox{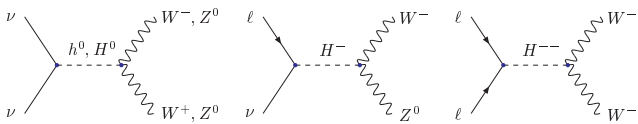}
\vspace*{0cm}
\caption{Additional diagrams that contribute to the amplitudes in 
Eqs.~(\ref{WW})--(\ref{SS}) when the Majorana neutrino acquires its
mass via a coupling to an SU(2)$_L$-triplet Higgs field.}
\label{fig:fig3}
\end{center}
\end{figure}

\newpage

Neutrino oscillation experiments do not measure the neutrino mass, but 
rather the absolute value of the mass-squared difference of two species 
of neutrinos,
$\Delta m^2$.  This implies a lower bound of $m_\nu \ge \sqrt{\Delta m^2}$
on the mass of one of the two participating neutrino species.  
Using Eq.~(\ref{BOUND}), one finds the upper bounds on the scale 
$\Lambda_{Maj}$
given in Table~1 for a variety of neutrino oscillation experiments.
These upper bounds are all within a few orders of magnitude of the Planck 
scale, $G_N^{-1/2} \simeq 1.2\times 10^{19}$ GeV, which is the 
scale before which quantum gravity must become relevant.  However, only the 
vacuum-oscillation interpretation of the solar neutrino deficit yields a 
scale that could be as large as the Planck scale.   In all other cases, 
we find that the physics of Majorana-neutrino mass generation must be below
the Planck scale.  Thus, if these neutrino masses arise from quantum gravity,
then the scale of quantum gravity must be somewhat less than the Planck scale.

The upper bounds on $\Lambda_{Maj}$ are also within a few orders of
magnitude of the grand-unification scale, ${\cal O}(10^{16})$ GeV. 
The LSND (Liquid Scintillator Neutrino Detector) and
atmospheric neutrino experiments yield an upper bound on $\Lambda_{Maj}$
slightly below the grand-unification scale, but the scale of Majorana-neutrino
mass generation could be less than the unification scale 
in a grand-unified model.  
For example, in a grand-unified model that makes use of the see-saw 
mechanism, the mass of the heavy Majorana neutrino $N$ could be equal to
a small Yukawa coupling times the vacuum-expectation value of the Higgs 
field that breaks the grand-unified group.

In this paper we have derived a model-independent upper bound on the scale
of Majorana-neutrino mass generation, Eq.~(\ref{BOUND}).  The upper bounds
on this scale implied by a variety of neutrino oscillation experiments
are listed in Table~1.  All but one of these bounds are less than
the Planck scale, and they are all within a few orders of 
magnitude of the grand-unification scale.

\begin{table}[t]
\caption{Neutrino mass-squared differences from a variety of neutrino
oscillation experiments, and their interpretations.  The last column gives
the upper bound on the scale of Majorana-neutrino mass generation,
Eq.~(\ref{BOUND}), for each interpretation.  
Table adapted from Ref.~[1].}
\begin{tabular}[4]{lllc}
Experiment&Favored Channel&$\Delta m^2$ ($\rm eV^2$)
&$\Lambda_{Maj}$ $({\rm GeV})$ $<$\\
\noalign{\vskip2pt}\hline\noalign{\vskip2pt}
LSND &$\bar\nu_\mu\to\bar\nu_e$&$0.2-2.0$&$9.8\times10^{14}$\\
Atmospheric&$\nu_\mu\to\nu_\tau$&$3.5\times10^{-3}$&$7.4\times10^{15}$\\
 Solar\\
\quad MSW (large angle)
&$\nu_e\to\nu_\mu$ or $\nu_\tau$&
$(1.3-18)\times10^{-5}$&$1.2\times10^{17}$\\
\quad MSW (small angle)&$\nu_e\to$ anything &$(0.4-1)\times10^{-5}$
&$2.2\times 10^{17}$\\
\quad Vacuum&$\nu_e\to\nu_\mu$ or
$\nu_\tau$ &$(0.05-5)\times10^{-10}$&$2.0\times10^{20}$\\
\end{tabular}
\label{massdiffs}
\end{table}

\section*{Acknowledgments}
\indent 
We are grateful for conversations with K.~Babu and D.~Dicus. 
This work was supported in part by the U.~S.~Department 
of Energy under contract No.~DOE~DE-FG02-91ER40677.  
We gratefully acknowledge the support of GAANN, under Grant No.~DE-P200A980724,
from the U.~S.~Department of Education for J.~M.~N.

\end{document}